\documentclass[aps,prl,twocolumn,groupedaddress,showpacs,superscriptaddress]{revtex4}
\usepackage{graphicx}

\begin{document}

\title{Pure Mott phases in confined ultra-cold atomic systems}

\author{V.G.~Rousseau}
\affiliation{Department of Physics and Astronomy, Louisiana State
  University, Baton Rouge, Louisiana 70803, USA}
\author{G.G.~Batrouni}
\affiliation{INLN, Universit\'e de Nice-Sophia Antipolis, CNRS; 
1361 route des Lucioles, 06560 Valbonne, France}
\affiliation{Centre for Quantum Technologies, National University of
  Singapore; 2 Science Drive 3 Singapore 117542}
\author{D.E.~Sheehy}
\author{J.~Moreno}
\author{M.~Jarrell}
\affiliation{Department of Physics and Astronomy, Louisiana State
  University, Baton Rouge, Louisiana 70803, USA} 

\begin{abstract}
We propose a novel scheme for confining atoms to optical lattices by
engineering a spatially-inhomogeneous hopping matrix
element in the  Hubbard-model (HM) description, a situation we term
off-diagonal confinement (ODC).  We show, via an exact numerical
solution of the boson HM with ODC, that this scheme possesses
distinct advantages over the conventional method of confining atoms
using an additional trapping potential, including 
incompressible Mott phases at commensurate filling and a phase diagram
that is similar to the uniform HM. The experimental implementation
of ODC will thus allow a more faithful realization of correlated
phases in cold atom experiments.
\end{abstract}

\maketitle

In recent years there has been a burgeoning
interest in the boson Hubbard (BH) model~\cite{Fisher89,Batrouni1990},
spurred by the expectation~\cite{Jaksch98} that modern laser cooling
and trapping techniques lead to the realization of this model in
cold atom experiments.  Several experimental
groups~\cite{Greiner02,Stoferle2004,Foelling,
  Campbell,Spielman07,Guarrera,Gemelke,Trotzky09} have achieved the BH
model with bosonic atoms in optical lattices, observing evidence of
the predicted Mott insulating and superfluid phases. Experimental 
probes include the condensate fraction and 
momentum distribution \cite{Spielman07}, noise correlations \cite{Guarrera} and 
RF spectroscopy \cite{Campbell}. 

\begin{figure}[h]
  \centerline{\includegraphics[width=0.40\textwidth]{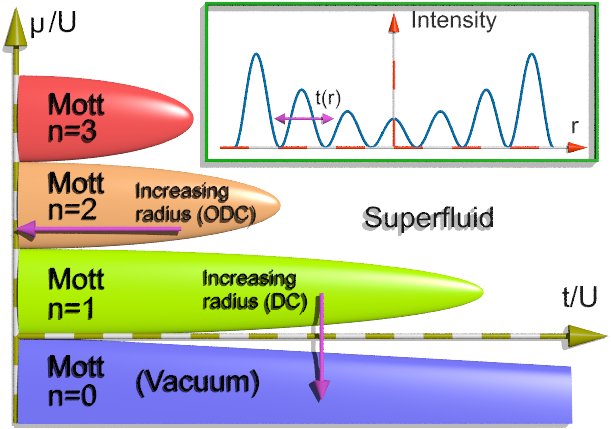}}
  \caption
    { (Color online) Schematic ground state phase diagram of the uniform boson Hubbard (BH)
      model~\cite{Fisher89}. The arrows show two possible sequences of local phases encountered in an atomic system
      as the radius $r$  from the center of the lattice  increases.
      Diagonal confinement (DC) is created by a decreasing chemical potential $\mu$, leading to a sequence of Mott and superfluid phases.
      Off-diagonal confinement (ODC) is created by a decreasing hopping parameter $t$, and can
      maintain the whole system in the same Mott lobe.
      Inset: A decreasing hopping parameter $t$ can be obtained by increasing the intensity of the optical lattice as one moves away from the center.}
  \label{SchematicPhaseDiagram}
\end{figure}

The ground state phase diagram for the uniform BH model with chemical
potential $\mu$, onsite repulsion $U$, and hopping parameter $t$ is
well known (Fig.~\ref{SchematicPhaseDiagram}), and consists of lobes of incompressible Mott insulating
phase (at commensurate fillings) surrounded by a regime of superfluid
order, with these states separated by quantum phase transitions~\cite{Fisher89,Batrouni1990}.
However, present-day cold atom experiments always occur in the
nonuniform environment of a smoothly-varying harmonic trap that
confines the atoms to the optical lattice, so that  
the actual Hamiltonian consists of the
BH model plus a parabolic single-particle potential.  Early
theoretical work~\cite{Batrouni2002} showed that this kind of diagonal
confinement (DC) does not entirely reproduce the physics of the BH
model. In particular, the model with DC is characterized by the
absence of energy gaps at commensurate fillings, as opposed to the BH
model. This results in the absence of true incompressible Mott
insulating phases and of a real  superfluid to Mott
transition in present-day experiments.

The physical origin of the lack of incompressible Mott phases in the
BH model with DC~\cite{Jaksch98,Batrouni2002} can be understood
within the local density approximation (LDA). According to the LDA,
the harmonic DC potential is treated by replacing $\mu \to \mu -
\frac{1}{2} m\Omega^2 r^2$ with $r$ the distance from the center of the lattice, $\Omega$ the
trapping frequency and $m$ the atom mass.  With increasing radius,
then, the system traverses the uniform-case phase diagram at fixed $t$ and a 
spatially-varying $\mu$ (vertical arrow on Fig.~\ref{SchematicPhaseDiagram}). 
Then, even if a system is locally in a Mott phase at the center of the trap, it must
enter a superfluid phase with increasing radius (yielding the
well-known shell structure~\cite{Foelling,Campbell,DeMarco2005,Batrouni2002}), thereby being
globally compressible.
Recent work has used the LDA to infer the homogeneous BH phase diagram from the 
DC experimental results~\cite{Trotzky09} and also to place a given DC system 
on the phase diagram of the uniform BH case~\cite{Batrouni2008}.
Nevertheless, it is still of great interest to
establish truly incompressible Mott phases of bosons in cold atomic
gas experiments. 

In this Letter, we propose that truly incompressible Mott phases
can be achieved via off-diagonal
confinement (ODC) of the bosons where the cloud is confined via a
spatially varying tunneling matrix element $t(r)$, which vanishes at the
boundary of the system. This can be realized, for example, by having the optical
lattice depth increase with increasing radius, as illustrated in the inset 
of Fig.~\ref{SchematicPhaseDiagram}. 
Although a system with ODC is still
spatially inhomogeneous, it can exhibit a true gapped and
incompressible Mott phase at commensurate fillings.
This can be easily understood by extending the
above LDA argument to a spatially-varying $t(r)$ instead of $\mu(r)$.  
Thus, within the LDA, a vanishing $t(r)$ with increasing $r$ corresponds to traversing
the uniform-case phase diagram at {\it fixed} chemical potential.  In
particular, a system that is locally in a Mott phase at the center of
the lattice will remain in the same Mott phase for all radii up to the
boundary where $t(r)$ becomes zero.  

Thus the ODC setup reproduces important aspects of the BH model, namely the presence of gaps and
true Mott phases at commensurate fillings, as we show via an exact
numerical calculation  for a one-dimensional model.
In addition, superfluid phases with ODC are found to lead to condensates (quasicondensates in
one dimension) with greater population than the model with DC, and
therefore can realize {\it clean} condensates.

Let us briefly discuss 
the experimental viability of achieving off-diagonal confinement via 
a spatially-dependent $t$.  A uniform periodic lattice of cold atoms
yields  a tunneling matrix element~\cite{BlochReview}
\begin{equation}
t\simeq \frac{4}{\sqrt{\pi}}E_r 
\bigg(\frac{V_0}{E_r}\bigg)^{\frac{3}{4}}\exp\bigg(-2\sqrt{\frac{V_0}{E_r}}\bigg),
\label{eq:tofv}
\end{equation}
with $E_r$ the recoil energy ($\displaystyle E_r = \frac{\hbar^2k^2}{2m}$, 
$k$ the optical lattice wave vector,  $m$ the atom mass) and $V_0$ the 
optical lattice depth.  
The exponential sensitivity of $t$ on $V_0$ implies that
even a modest increase in $V_0$ with increasing radius will yield a rapid suppression
of $t$, confining the cloud to the optical lattice.
Taking parameters from Ref.~\onlinecite{Greiner02}, which has
$E_r \simeq 2.13\times 10^{-13} J$, this sensitivity is illustrated by noting that 
the hopping time $\tau\sim \hbar/t$ changes by  two orders of magnitude as $V_0/E_r$ is varied 
(from $0.3$ms at $V_0/E_r=3$ to 
 $26$ms at  $V_0/E_r=22$).   

Although conventional optical lattices 
created with interfering laser beams typically possess a uniform $t$,
recently the Greiner group has developed a novel holographic method
that can produce an arbitrary optical potential using a mask~\cite{Waseem}.
An optical lattice that produces a spatially-varying $t$ will also yield
a spatially-varying Coulomb repulsion $U$.  However, examining the expression for this quantity~\cite{BlochReview} 
$U = \sqrt{8/\pi} ka E_r (V_0/E_r)^{3/4}$, with $a$ the atom scattering length, we note that $U$ 
is only power-law sensitive to $V_0$. Therefore, 
a spatially-varying $V_0$ will lead to relatively small variations of $U$ and  larger 
changes in $t$.
Moreover, variations of $U$ can be taken into account by generalizing the LDA picture to a varying 
$t/U$, yielding same conclusions.
Thus, we expect that such off-diagonal confinement of atoms in an optical lattice will be experimentally realizable in the 
near future, adding to the rapidly-growing cold-atom toolbox~\cite{JakschToolbox}.

The above LDA argument suggests that
experiments using ODC will achieve a true Mott phase for any
spatially-varying $t(r)$.  
To illustrate this,
we consider the following one-dimensional model:
\begin{equation}
 \label{eq:Ham}
 \hat\mathcal H=-\sum_{\big\langle ij\big\rangle} t_{ij}\big(a_i^\dagger a_j+a_j^\dagger a_i\big)+\frac{U}{2}\sum_i\hat n_i\big(\hat n_i-1\big) 
\end{equation}

\begin{figure}[h]
  \centerline{\includegraphics[width=0.40\textwidth]{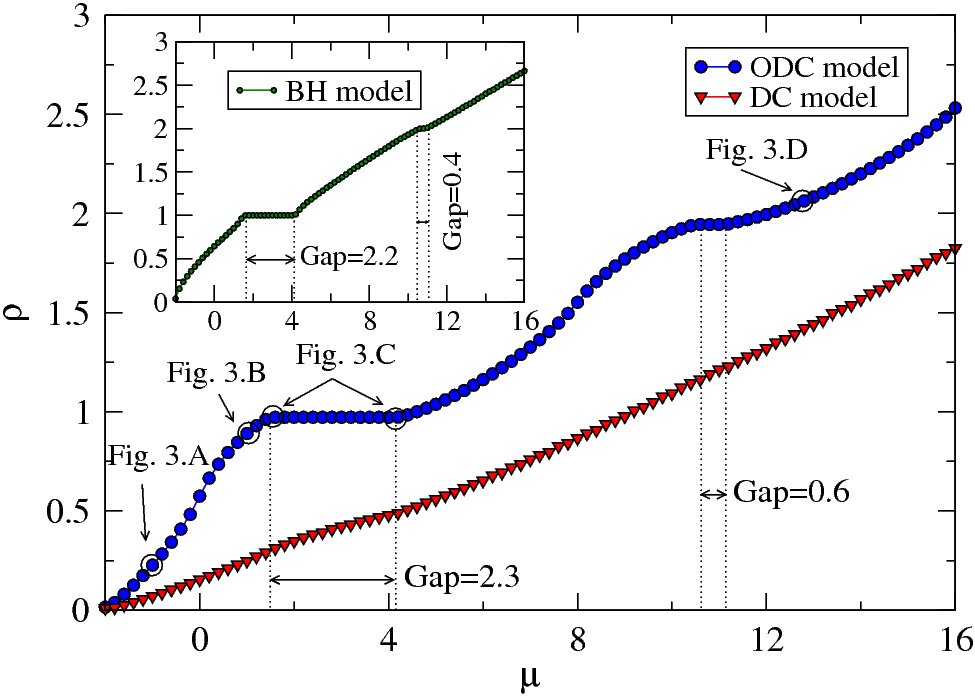}}
  \caption
    { (Color online) The density as a function of the chemical
      potential, for $U=8$. The ODC model exhibits Mott gaps at
      commensurate fillings as in the periodic case (inset). Results
      for the DC model with a trapping parameter $W=0.008$ are shown
      for comparison.  }
  \label{RhoVsMu}
\end{figure}
The sum over $i$ runs in the range $\big[0;L-1\big]$, with $L$ the number of lattice sites, and the sum over $\big\langle ij\big\rangle$ is restricted to pairs of
first neighbors.
The bosonic operators $a_i^\dagger$ and $a_i^{\phantom\dagger}$ create and
annihilate a particle on site $i$, respectively,  and
$\hat n_i=a_i^\dagger a_i^{\phantom\dagger}$. 
The local hopping parameter $t_{ij}$ follows an inverted parabola, \mbox{$t_{ij}=t(i+j+1)(2L-i-j-1)/L^2$},
with a maximum value of $t$ at the center of the lattice, $t_{L/2-1,L/2}=t$, and a vanishing
value at the edges ($t_{-1,0}=t_{L-1,L}=0$). Finally, we choose $t=1$ in order to set the
energy scale.

We solve the model exactly by means of quantum Monte Carlo
simulations, by making use of the Stochastic Green Function (SGF)
algorithm~\cite{SGF} with directed updates~\cite{DirectedSGF}.  The SGF
algorithm can be implemented in the canonical ensemble  as well as
in the grand-canonical ensemble. In the following, we take
advantage of this feature. When working in the  canonical ensemble, the number of
particles $N$ is chosen and remains strictly constant during the entire
simulation. In the grand-canonical ensemble, we add the usual term 
$-\mu\hat N$ to the Hamiltonian, with $\hat N=\sum_i\hat n_i$,
in order to control the mean number of particles $N=\big\langle\hat N\big\rangle$
via the chemical potential $\mu$.
In the following we consider a lattice with $L=70$ sites, since this
is the typical size currently accessible in 
one-dimensional optical lattice experiments. We have used increasing values of the
inverse temperature $\beta$, and found that the ground-state properties are captured
with $\beta=20$. All presented results correspond to this inverse temperature.
We begin by investigating the
behavior of the density $\rho=N/L$ as a function of the chemical potential
$\mu$. Here it is convenient to make use of the grand-canonical ensemble, since the
chemical potential is the control parameter.  
Fig.~\ref{RhoVsMu} displays results for  $U=8$.

\begin{figure}[h]
  \centerline{\includegraphics[width=0.40\textwidth]{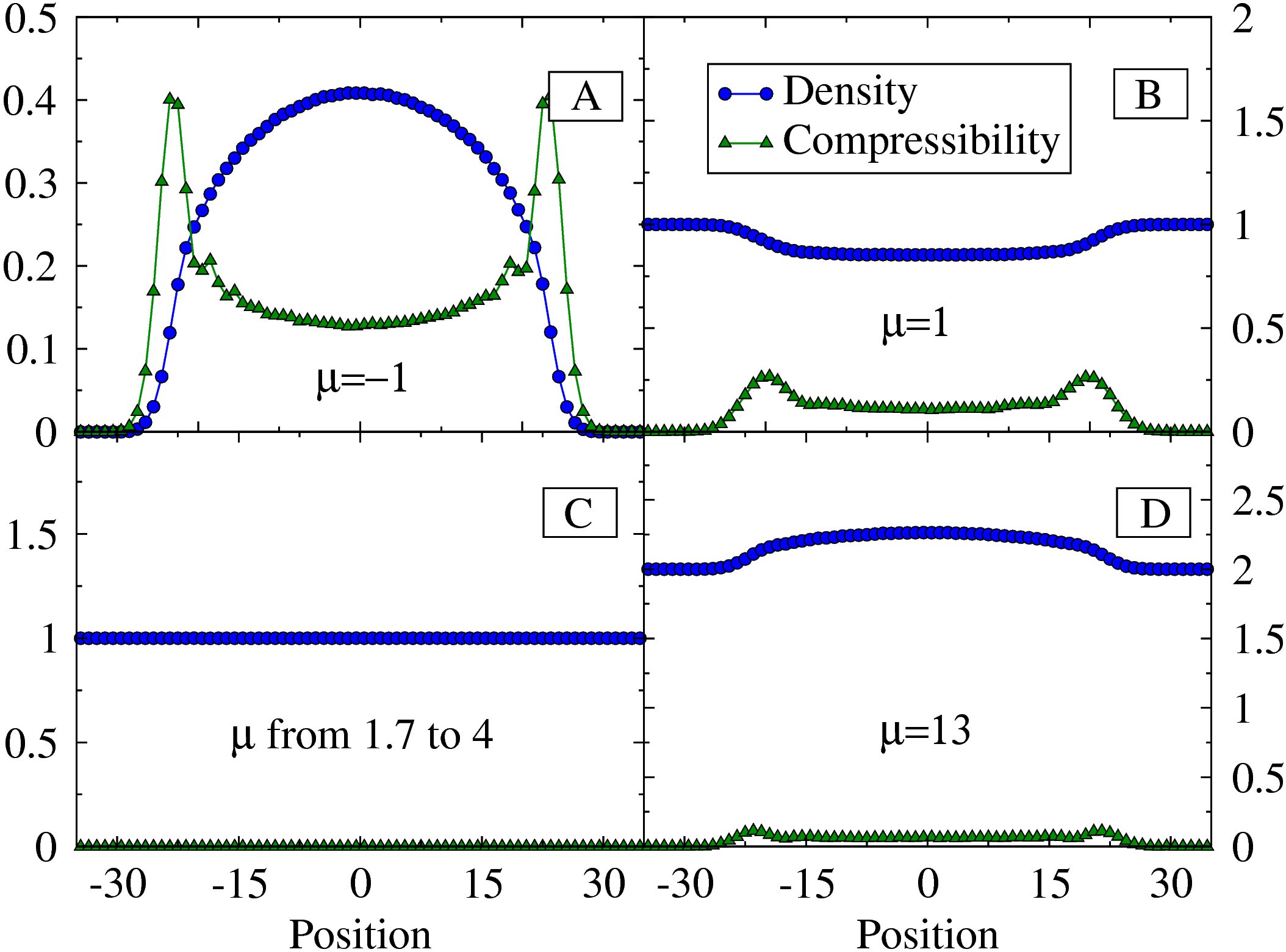}}
  \caption
    {
      (Color online) The local density and local compressibility, for $U=8$. A superfluid cloud grows in the center of the lattice,
      as the number of particles increases (Panel A). Contrary to the DC model, Mott regions in the ODC model
      form at the edges of the lattice (Panel B), then extend towards the center. At commensurate filling (Panel C),
      the phase is a pure incompressible Mott insulator. Adding particles above commensurate filling breaks the Mott plateau in the center,
      which becomes locally superfluid (Panel D). 
    }
  \label{DensityProfiles}
\end{figure}

A comparison between the results in the main panel and the inset in Fig.~\ref{RhoVsMu}
shows that while the usual model with DC  does not display any
incompressible phase, the model with ODC has pure Mott gaps at
commensurate fillings as in the homogeneous case. 
For these fillings, the number of particles per site sticks to integer values
and is constant over the whole lattice. Note also that the gaps of the ODC model
and homogeneous BH model are similar in size.
The results for the DC model
correspond to a parabolic potential $W\sum_{i}(i-L/2)^2\hat n_i$ with
$W=0.008$. For this case, the size of the lattice has been increased
to $L=100$ in order to avoid an overflow of the particles. Note that
such an overflow never occurs in the ODC model, since the local
hopping parameter vanishes at the edges of the
lattice.

Fig.~\ref{DensityProfiles} shows typical density profiles and
the associated local compressibility, defined as the response of the local number of particles
to a change of the chemical potential ~\cite{Wessel},
$\displaystyle \kappa_i=\frac{\partial\langle\hat n_i\rangle}{\partial\mu}=
  \int_0^\beta\big[\big\langle \hat n_i(\tau)\hat N\big\rangle-\big\langle\hat n_i(\tau)\big\rangle\big\langle\hat N\big\rangle\big]d\tau$.
For pure Mott phases this quantity %takes the useful form
vanishes at commensurate fillings for each lattice site $i$.

\begin{figure}[h]
  \centerline{\includegraphics[width=0.25\textwidth,angle=270]{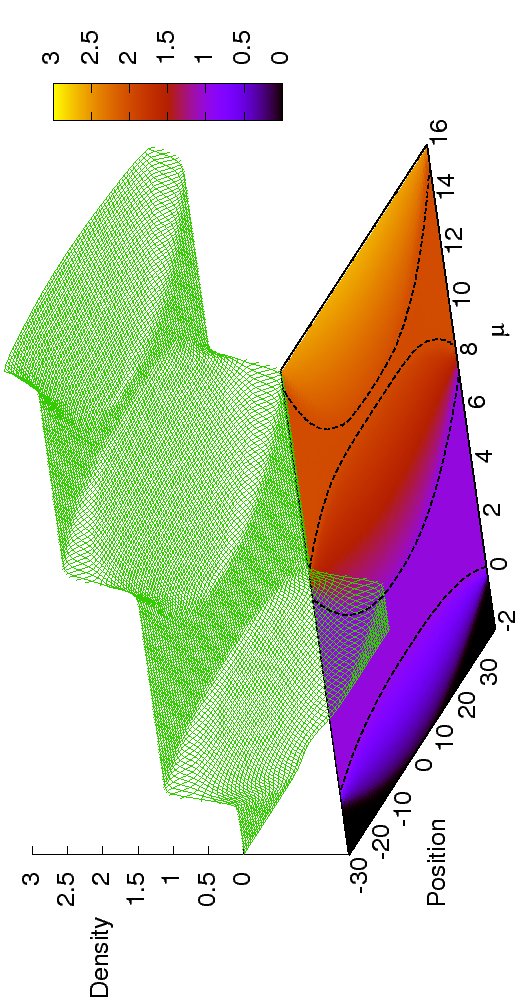}}
  \caption
    { (Color online) The density as a function of the position and the
      chemical potential for $U=8$. The Mott plateaux at $\rho=1$ and $\rho=2$ appear clearly, and extend over the whole lattice.}
  \label{Density3D}
\end{figure}

The density profiles in 
Fig.~\ref{DensityProfiles} can be qualitatively understood within the above LDA argument, in which
increasing radius corresponds to a decreasing $t$ and a horizontal path in the 
homogeneous BH phase  diagram of Fig.~\ref{SchematicPhaseDiagram}. 
Thus, Fig.~\ref{DensityProfiles}A shows a system that is locally
superfluid in the center, and enters the vacuum as the radius increases and 
the ratio $t/U$ decreases. For greater $\mu$, Fig.~\ref{DensityProfiles}B
corresponds to a path beginning in the superfluid phase and crossing into the $n=1$ Mott phase of 
Fig.~\ref{SchematicPhaseDiagram}, therefore the system 
is still globally compressible with a local Mott
phase at the edge.   The incompressible Mott phase in 
Fig.~\ref{DensityProfiles}C corresponds to a LDA path that remains entirely inside one Mott lobe.
Fig.~\ref{DensityProfiles}D at higher $\mu$ is qualitatively
similar to Fig.~\ref{DensityProfiles}B, but with the inner superfluid having a higher local density than the Mott region at the edges. 
Finally, all density profiles 
corresponding to the $\mu$ values explored in 
Fig.~\ref{RhoVsMu} are represented on Fig.~\ref{Density3D}. The
figure clearly shows the Mott plateaux at $\rho=1$ and $\rho=2$
extending over the whole lattice.

\begin{figure}[h]
  \centerline{\includegraphics[width=0.40\textwidth]{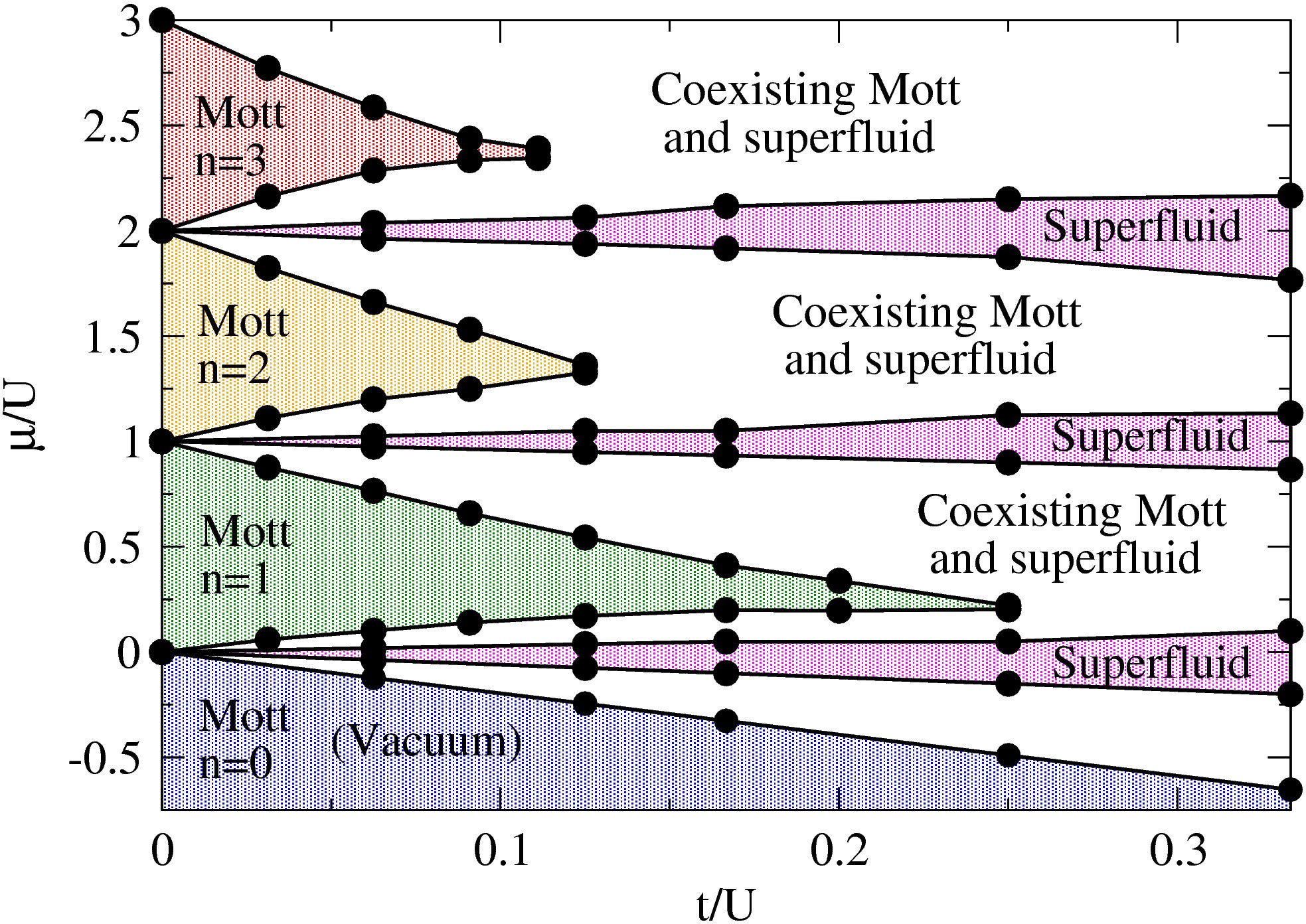}}
  \caption
    {
      (Color online) The ground state phase diagram of the ODC model given by 
Eq.~(\ref{eq:Ham}),
showing lobes of incompressible pure Mott phase (like Fig.~\ref{DensityProfiles}C), 
and regions of compressible superfluid and compressible
coexisting Mott and superfluid phases (like Figs.~\ref{DensityProfiles}A, \ref{DensityProfiles}B, and \ref{DensityProfiles}D). 
    }
  \label{PhaseDiagram}
\end{figure}

Next, we work in the canonical ensemble at fixed densities
and scan the values of the onsite repulsion
parameter $U$ for which the system exhibits the Mott insulating
state. In the canonical ensemble the ground state chemical potential is not a control parameter,
but instead it is computed as $\mu(N)=E(N+1)-E(N)$. 
Fig.~\ref{PhaseDiagram} shows the phase diagram which exhibits pure
Mott lobes similar to those of the homogeneous BH model~\cite{Batrouni1990}, 
which again %we emphasize that such pure Mott lobes 
are not present in the DC model~\cite{Batrouni2002}. In
addition, new lobes where the phase is purely superfluid arise as the
onsite interaction decreases. Another particularity of the ODC model
is that coexistence of Mott and superfluid can occur at any value of
the onsite repulsion $U$. This mixed state is absent in the uniform BH
model; and in the DC model there is a critical value of $U$~\cite{Rigol} 
below which the system is purely superfluid. 

\begin{figure}[h]
  \centerline{\includegraphics[width=0.45\textwidth]{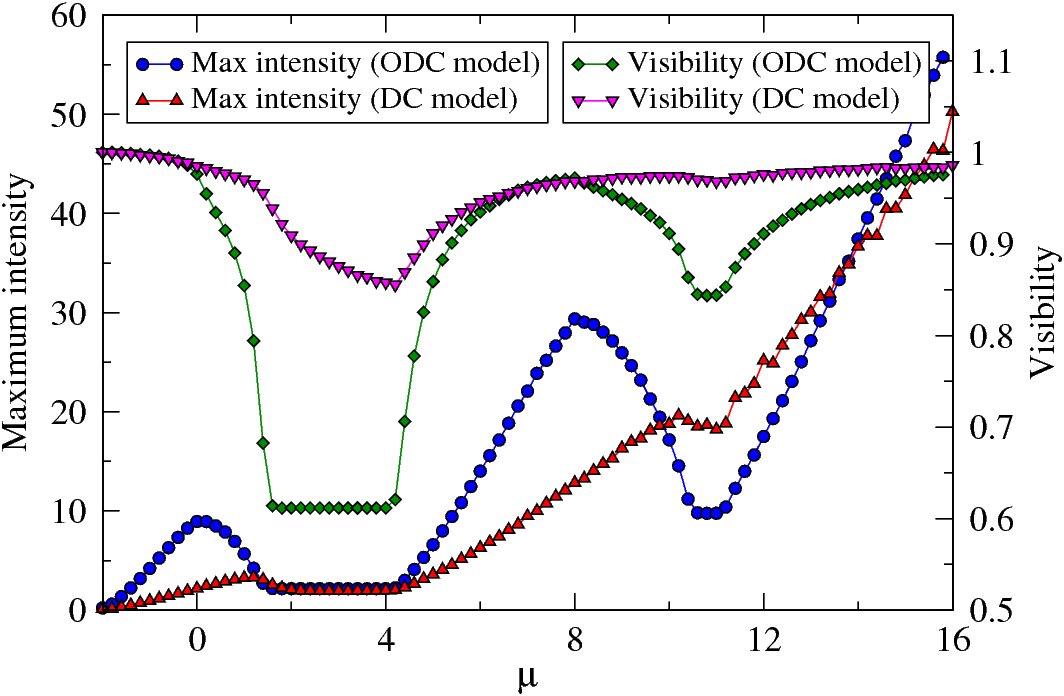}}
  \caption
    {(Color online) The maximum intensity of the interference pattern
      and the visibility as functions of the chemical potential, for $U=8$.}
  \label{Visibility}
\end{figure}

Finally, we make connection with an experimental probe by plotting in
Fig.~\ref{Visibility} the maximum intensity $I_{max}$ of the
interference pattern (the population of the quasi-condensate) as a
function of the chemical potential. The visibility 
$\mathcal V =(I_{max}-I_{min})/(I_{max}+I_{min})$ is
also shown. The variations in the visibility are stronger with ODC than with DC, 
since in the former case we are in the
presence of a true superfluid-to-Mott transition. 
The maximum intensity curve also reveals another
interesting feature of the ODC model: condensates have greater
population than the DC model. Thus, the experimental realization of
the ODC model should improve the quality of condensates.

In this Letter, we proposed a novel scheme for confining atoms on 
optical lattices based
on off-diagonal confinement, in which the Bose-Hubbard tunneling matrix
element is engineered to vanish at the system boundary.
We show by using an exact quantum Monte Carlo method that our model reproduces 
important aspects of the boson Hubbard model.  In particular, truly
incompressible bosonic Mott phases occur at commensurate fillings
and strong interactions, an important
requirement for quantum-computing~\cite{JakschToolbox}. Indeed, optical lattices provide intrinsically
scalable and well-defined sets of qubits. These qubits must be initialized in a "zero" state,
which corresponds to having one particle per site. This is precisely achieved by having the system 
in a pure Mott phase. Additionally
the establishment of a global Mott phase will pave the way to further
experimental simulations of the Mott-superfluid transition, a problem of
broad interest due to its connections to condensed-matter systems such as the
high-temperature superconductors~\cite{LeeNagaosa}.
In addition, our simulations show that the model leads to
highly populated condensates, which are of great interest in the
production of coherent matter waves.

% The original new paragraph of Dan

%To achieve the ODC, we can assume it may be easiest to start with a conventional
%DC cloud (a system that has already been taken to low temperatures) and 
%adiabatically apply the ODC optical potential.  
%We have studied the entropy of the DC and ODC phases in a hardcore-boson model
%at finite temperatures, finding, for comparable system size, a comparable entropy 
%in the ODC and DC states, implying that a low temperature can be maintained upon adiabatically
%moving from DC to ODC. 
%Additionally, one may ask how  equilibrium is established with ODC given that the local hopping time $\tau\sim \hbar/t$ is spatially varying;
%rapid equilibration may be stymied if much of the system possesses a large value of $\tau$. To 
%address this it may be necessary to engineer ODC to have $V_0$ increase rapidly over a few lattice
%sites near the edge.  Given the abovementioned rapid variation of $\tau$ with $V_0$, this should be 
%feasible.

% The parapraph after Val's modifications

To achieve the ODC, it may be easiest to start with a conventional DC cloud (a
system that has already been taken to low temperatures) and adiabatically apply
the ODC optical potential. We have studied the entropy of the DC and ODC phases
in a hardcore-boson model at finite temperatures, finding, for comparable system
size, a comparable entropy in the ODC and DC states, implying that a low
temperature can be maintained upon adiabatically moving from DC to ODC.
Additionally, one may ask how  equilibrium is established with ODC given that the local hopping time $\tau\sim \hbar/t$ is spatially varying;
rapid equilibration may be stymied if much of the system possesses a large value of $\tau$. To 
address this it may be necessary to engineer ODC to have $V_0$ increase rapidly over a few lattice
sites near the edge.  Given the abovementioned rapid variation of $\tau$ with $V_0$, this should be 
feasible.

The flexibility of the techniques used in present-day
experiments allows the realization of our model,
and should open new perspectives on the physics of Mott phases.
Future work will utilize the 
Stochastic Green Function (SGF) algorithm~\cite{SGF,DirectedSGF} to 
compute other experimentally relevant quantities such as the
momentum distribution~\cite{Spielman07}, noise correlations~\cite{Guarrera} and 
RF spectroscopies~\cite{Campbell}.  
Finally, we note that our model can also be easily generalized 
to fermionic systems where we expect 
a similarly rich behavior. 

\begin{acknowledgments}
We express special thanks to M.~Greiner for useful discussions about experimental 
setups. 
This work was supported by NSF OISE-0952300, and by Louisiana Board of Regents grant 
 LEQSF (2008-11)-RD-A-10. We are also grateful to U.~Sam for financial support.
\end{acknowledgments}

\end{document}